\overfullrule 0pt
\sequentialequations

 \pubnum{6087T}
 \date{December 20, 1994}
 \pubtype{(T/E)}
 \titlepage

 \title{QCD CONSTRAINTS ON THE SHAPE OF
POLARIZED QUARK AND GLUON DISTRIBUTIONS\foot{%
Work supported in part by Department of Energy contracts
DE--AC03--76SF00515 and DE--AC02--76ER03069,
and by Fondo    Nacional de Investigaci\'on Cient\'\i fica
   y Tecnol\'ogica, Chile, contract 1931120. }}
{\baselineskip 17pt
 \author{Stanley J. Brodsky}
 \SLAC
 \andauthor{Matthias Burkardt\foot{Present address:
Institute for Nuclear Theory, University of Washington, Seattle,
Washington 98195.}}
 \address{Center for Theoretical Physics, Laboratory for
Nuclear Science \break
and Department of Physics, Massachusetts Institute of Technology,
Cambridge, Massachusetts 02139}
 \andauthor{Ivan Schmidt}
\address{Universidad Federico Santa Mar\'\i a \break
Casilla 110--V, Valpara\'\i so, Chile }
\par}
\vfill
\submit{Nuclear Physics B.}
\vfill
\endpage

\vbox{\vskip 1.5in}

\centerline{ABSTRACT}
We develop simple analytic representations of the polarized
quark and gluon distributions in the nucleon at low $Q^2$
which incorporate general
constraints obtained from the requirements of color coherence
of gluon couplings at $x \sim 0$ and the helicity retention properties of
perturbative QCD couplings at $x \sim 1.$ The unpolarized predictions
are similar to the $D_0'$ distributions given by Martin,
Roberts, and  Stirling.
The predictions for the quark helicity distributions
are compared with
polarized structure functions
measured by the E142 experiment at SLAC
and the SMC experiment at CERN.

\vfill
\endpage

{\bf \chapter {INTRODUCTION}}

\REF\SMC{B. Adeva, \etal, Phys. Lett. {\bf B302}, 533 (1993).}

\REF\Emlyn{P. L. Anthony, \etal, Phys. Rev. Lett. {\bf 71}, 959 (1993).
K. Abe, \etal, SLAC-PUB-6508, (1994).
For recent
reviews see R. G. Milner, Plenary Talk at SPIN-94,
MIT preprint LNS 94-88 (1994) and
S. Forte, CERN-TH.7453/94, to be published in the
proceedings of the Tennessee International Symposium on Radiative
Corrections, Gatlinburg (1994).}

\REF\SMCp{D. Adams \etal, Phys. Lett. {\bf B329}, 399 (1994).}

\REF\GRV{M. Gluck, E. Reya, and A. Vogt,
Phys. Lett. {\bf B306} 391, (1993) and references, therein. }

\REF\Dick{R. Blankenbecler and S. J. Brodsky, Phys. Rev {\bf D10},
2973 (1974); J.~F.~Gunion, Phys. Rev. {\bf D10}, 242 (1974).
S. J. Brodsky and G. P. Lepage, Proceedings of the
1979 Summer Institute on Particle Physics, SLAC, (1979).}

Measurements of polarization correlations in high momentum
transfer reactions can provide highly
sensitive tests  of the underlying structure and
dynamics of hadrons.
The most direct information on
the light-cone momentum
distributions of helicity-aligned and
helicity-anti-aligned quarks in nucleons is obtained from
deep inelastic scattering of polarized leptons on polarized
targets.  Recent fixed-target measurements, including
the CERN SMC muon-deuteron
experiment\refmark\SMC,
the electron-$He^3$ and electron-proton experiments E142 and E143
at SLAC\refmark\Emlyn,~ and the SMC muon-proton
experiment\refmark\SMCp~
are now providing important new constraints on the
proton and neutron  helicity-dependent structure functions.

Although the $Q^2$-evolution of structure functions is well-predicted by
perturbative QCD, the initial shape of these distributions reflects the
non-perturbative quark and gluon dynamics of the bound-state solutions of
QCD.  Nevertheless, it is possible to predict some aspects of
the shape of the input
nucleon structure functions from perturbative arguments
alone.  In this paper, we will develop simple analytic representations of
the quark and gluon helicity distributions which incorporate
general constraints  obtained from the
color
coherence of the
gluon couplings at $x \sim 0,$ and  the helicity structure of
perturbative QCD couplings at $x \sim 1.$
Since we work at the bound-state scale,
we can directly impose global sum rules and symmetries such as
the axial coupling constraint
$\Delta u =\Delta d = g_A,$
from neutron beta decay.
The parameterizations we use have
the minimal number of parameters needed to satisfy
all of the constraints.
The  predicted forms for the quark and gluon helicity distributions
$\Delta q(x)= q^+(x) - q^-(x)$ and
$\Delta G(x)= G^+(x) - G^-(x)$
should provide
useful guides to the expected shapes of the
polarized structure
functions and an understanding of how the helicity
content of the nucleon is distributed among its constituents.
Eventually these fundamental distributions should
be computable using non-perturbative methods such as
lattice gauge theory or light-cone
Hamiltonian diagonalization.

The
structure functions discussed in this paper
are meant to reflect the intrinsic
bound state structure of the nucleons, and thus they
strictly apply only at low resolution scales
$Q^2 < Q^2_0$ where QCD evolution can be neglected.
They can be used at large $Q^2$ as the input distributions for
perturbative QCD evolution, as in the analyses of Ref. [\GRV].
However, as we discuss below, the actual implementation
the evolution program must
take into account the fact
that for $x \sim 1$ the bound state quark which is struck
in deep lepton inelastic
scattering
is far off shell, thus suppressing its gluon radiation.

\REF\Farrar{G.  R.  Farrar and D.  R.  Jackson, Phys.  Rev.  Lett. {\bf
35}, 1416 (1975).}

\REF\LB{G.  P.  Lepage and S.  J.  Brodsky, Phys.  Rev. {\bf D22}, 2157
(1980).}

The polarized quark and gluon distributions $G_{q/H}(x,\lambda, Q)$
and $ G_{g/H}(x,\lambda, Q)$ of a hadron are most simply represented as
probability distributions  determined by the  light-cone wavefunctions
$\psi_n(x_i,k_{\perp i}, \lambda_i),$
where $\sum^n_{i=1} x_i = 1,$ and $\sum^n_{i=1}
k_{\perp_i} = 0_\perp.$
The square of the invariant mass of an n-particle Fock
State configuration in the wavefunction  is ${\cal M}_n^2 = \sum^n_{i=1}
(k^2_{\perp i} + m^2_i)/x_i.$
Thus the kinematical regime where one quark
has nearly all of the light-cone momentum $x \sim 1,$ and the remaining
constituents have $x_i \sim 0,$ represents
a very far off-shell configuration of
a bound state wavefunction.
In the limit $x \to 1,$ the Feynman virtuality of
the struck parton in a bound state becomes far off-shell and space-like:
$k_F^2-m^2= x(M^2_H- {\cal M}^2)  \to -\mu^2/(1-x),$ where $\mu$ is the
invariant mass of the system of stopped constituents.
If one assumes that the
bound state  wavefunction of the hadron is
dominated by the lowest invariant
mass partonic states, then the constituents can attain  far off-shell
configurations only by  exchanging  hard gluons;
thus the leading behavior at
large virtuality can be computed simply by iterating the gluon exchange
interaction kernel\refmark {\Dick, \Farrar, \LB}.
This conforms to the usual
ansatz of perturbative QCD that hard perturbative contributions dominate
amplitudes involving high momentum transfer compared to the contributions
arising from non-perturbative sources.

Thus, because of asymptotic freedom, the leading order contributions in
$\alpha_s(k_F^2)$ to the quark and gluon
distributions at $x \to 1$ can be
computed in perturbative QCD from minimally connected tree graphs.
For example,
in the case of the nucleon structure functions,
the dominant amplitude is
derived from graphs where the three valence quarks
exchange two hard gluons.
The tree amplitude is then convoluted with the
nucleon distribution amplitude
$\phi(x_i,k_F^2)$ which is obtained by integrating
the valence three-quark
nucleon wavefunction $\psi_3(x_i,k_{\perp i},
\lambda_i),$ over transverse
momenta up to the scale $k_F^2$\refmark\LB.
The $dk_\perp d\phi $ azimuthal
loop integrations project out only the $L_z=0$
component of the three-quark
nucleon wavefunction.  Thus, in amplitudes
controlled by the short distance
structure of the hadron's valence wavefunction,
orbital angular momentum can be
ignored, and the valence quark helicities sum to the hadron helicity.

\REF\Bj{J.  D.  Bjorken, Phys.  Rev. {\bf D1}, 1376 (1970).}

The limiting power-law behavior at $x \to 1$ of the helicity-dependent
distributions derived from the minimally-connected graphs is
$$G_{q/H} \sim (1-x)^p,$$ where $$p=2n-1+2\Delta S_z.$$ Here $n$ is the
minimal number of spectator quark lines, and $\Delta S_z = \vert
S^q_z-S^H_z \vert = 0$ or $1$ for parallel or anti-parallel quark and
proton helicities, respectively\refmark\Dick. This counting rule
reflects the fact that the valence Fock states with the minimum number of
constituents give the leading contribution to structure functions when
one quark carries nearly all of the light-cone momentum; just on
phase-space grounds alone, Fock states with a higher number of partons
must give structure functions which fall off faster at $x \to 1$. The
helicity dependence of the counting rule
also reflects the helicity retention
properties of the gauge couplings: a quark with a large momentum fraction
of the hadron also tends to carry its helicity.  The antiparallel
helicity quark is suppressed by a relative factor $(1-x)^2.$ Similarly,
in the case of a splitting function such as $q \to q g$ or $g \to \bar q
q,$ the sign of the helicity of the parent parton is transferred to the
constituent with the largest momentum fraction\refmark\Bj. The counting
rule for valence quarks can be combined with the splitting functions to
predict the $x \to 1$ behavior of gluon and non-valence quark
distributions.  In particular, the gluon distribution of non-exotic
hadrons must fall by at least one power faster than the respective quark
distributions.

\REF\BGD{E. Bloom and F.  Gilman, Phys.
Rev. Lett. {\bf 25}, 1140 (1970).}

\REF\Drell{S. D.  Drell and T.-M.Yan, Phys. Rev.
Lett.{\bf 24}, 181 (1970).}

\REF\Arnold{A. Lung, \etal,
Phys. Rev. Lett. {\bf 70}, 718 (1993).}

\REF\BLreview{%
S.  J.  Brodsky and  G. P.  Lepage,
in  {\it Perturbative Quantum Chromodynamics,}
Ed. by A. H.
Mueller,  World Scientific Publ. Co.(1989).}

\REF\Gribov{V.  N.  Gribov and L.  N.  Lipatov, Sov.  J.  Nucl.  Phys.
{\bf 15}, 438, 675 (1972).}

The counting rules for the end-point-behavior of  quark and gluon
helicity distributions can also be derived
from duality, \ie, continuity between
the physics of exclusive and inclusive channels at fixed invariant
mass\refmark\BGD.  As shown by Drell and
Yan\refmark\Drell, a quark structure
function $G_{q/H} \sim (1-x)^{2n-1}$ at $x \to 1$
if the corresponding form
factor $F(Q^2) \sim (1/Q^2)^n$ at large $Q^2$.
Recent measurements of elastic
electron-proton scattering at SLAC\refmark\Arnold~
are compatible with the
perturbative QCD predictions\refmark\BLreview~
for both the  helicity-conserving $F_1(Q^2)$  and
helicity-changing $F_2(Q^2)$ form factors:
$Q^4F_1(Q^2)$ and $Q^6F_2(Q^2)$
become approximately constant at large $Q^2$.  The  power-law fall-off of
the form factors  corresponds to the helicity-parallel
and helicity-antiparallel
quark distributions behaving  at $x \to 1$ as $(1-x)^3$ and $(1-x)^5,$
respectively, in agreement with the counting rules.
The leading exponent for
quark distributions is odd in the case of baryons and even for mesons in
agreement with the Gribov-Lipatov crossing rule\refmark\Gribov.

\REF\PDG{Particle Data Group,
Phys. Rev.  {\bf D45} (1992) .}

\REF\Jaffe{
J. Ellis and M. Karliner,
CERN-TH-7324/94,
Phys. Lett. {\bf B313}, 131 (1993). See also
R.  L.  Jaffe and A.
Manohar, Nucl.  Phys. {\bf B321}, 343 (1989);
F. E. Close, R. G. Roberts,
Phys. Lett. {\bf B316} 165,  (1993);
S.~D.~Bass, A. W. Thomas,
Cavendish preprint-HEP-93-4, (1993).}

\REF\Bourquin{M. Bourquin \etal, Z. Phys. {\bf C21}, 27 (1983).}

The counting rule predictions for the quark and gluon distributions  are
relevant at  low momentum transfer
scales $Q_0 \sim  \Lambda_{QCD}$ in which the controlling physics is
that of the  hadronic  bound state rather than the radiative corrections
associated with structure function evolution.
At the hadronic scale one can
normalize the non-singlet quark helicity
content of the proton and neutron using
the constraint from $\beta$ decay\refmark\PDG:
$$\Delta u - \Delta d = {g_A\over g_V} =
1.2573 \pm 0.0028\ .$$
where $\Delta q_i(x) =  q^+_i(x) - q^-_i(x)$ with $i=u,d,s$
is the difference of the helicity-aligned and helicity-anti-aligned
quark distributions  in the proton,
and $\Delta q_i = \int_0^1 dx \Delta q_i(x)$ is the integrated moment.
[In the standard notation
$q^+(x,Q)=G_{q/p}(x,\lambda_q=\lambda_p,Q)+G_{\bar
q/p}(x,\bar\lambda_q=\lambda_p,Q)$ so that
both quark and anti-quark contributions
are included.]  In addition, if one assumes $SU(3)$ flavor symmetry,
hyperon decay also implies a polarized strange quark
component in the proton wavefunction\refmark{\Jaffe,~\Bourquin}
$${\Delta u + \Delta d -2 \Delta s\over \sqrt 3} = 0.39\ .$$
Thus only one normalization is left undetermined.

The presence of polarized gluons in the nucleon wavefunction implies that
polarized strange quarks  contribute to the nucleon  helicity-dependent
structure functions at some level.
There is also evidence from neutrino-proton
elastic scattering that the proton has a
significant polarized strange quark content\refmark\Jaffe.
Our parameterization of the polarized strange quark
distributions with a significant helicity fraction
$\Delta s \simeq 0.10$ and a small momentum fraction
$\VEV {x_s} \simeq 0.035.$ The shape represents
the sum of  contributions from radiatively-generated
$s$ + $\bar s$ quarks as well as intrinsic strange quarks
intrinsic to the nucleon bound states.

\REF\rad{%
For recent analyses of the radiative corrections to the Bjorken sum rule
see J.~Ellis and M. Karliner,
Ref. [\Jaffe], and S. J.
Brodsky and H. J. Lu, SLAC-PUB-6481 (1994)
to be published in Phys. Rev. D.}

\REF\BergerQiu{An analysis of the
evolution of the  helicity-dependent
quark and gluon structure functions is given in
E. L. Berger and J. Qiu, Phys. Rev. {\bf D40}, 3128 (1989).}

The helicity-dependent structure function $g_1(x,Q^2)$ measured in deep
inelastic polarized-lepton polarized-proton  scattering  can
be identified in the Bjorken scaling region with the quark helicity
asymmetry:
$$g_1(x,Q^2) = \half \sum_i e_q^2\Delta q(x,Q^2)\ .$$
The first moment of the proton-neutron difference has zero
anomalous dimension and
satisfies the Bjorken sum rule:\refmark\Bj
$$\int^1_0 dx [g^p_1(x,Q^2) - g^n_1(x,Q^2)]=
 {1\over 6}\ {g_A \over g_V}
 \left(1-{\alpha_s(Q^2)\over \pi}  + \cdots\right)\ ,$$
where the last factor represents the radiative corrections from
hard gluon interactions in the electron-quark scattering
process\refmark\rad.
Thus the QCD radiative corrections\refmark\BergerQiu~
to the helicity-dependent structure functions can modify the shape of the
distributions, within the global constraint of
the Bjorken sum rule.

\REF\SS{For further discussion,
see S.  J.  Brodsky and  G. P.  Lepage,  Ref. [\Dick].}

At high $Q^2,$
the radiation from the struck quark line
increases the effective
power law fall-off
$(1-x)^p$ of  structure functions relative to the underlying quark
distributions:
$\Delta p = (4 C_F/\beta_1) \log (\log Q^2/\Lambda^2)/
(\log Q_0^2/\Lambda^2)$  where $C_F= 4/3$
and $\beta_1=11-(2/3) n_f.$
The counting rule predictions
for the power $p$ thus provide a
{\it lower bound}\
for the effective exponent of quark structure functions  at
high $Q^2 > Q^2_0$. However,
in the end-point region $x \sim 1,$ the struck quark
is far off-shell and the radiation is quenched
since one cannot evolve $Q^2$
below $Q^2_0 \simeq k^2_f= -(\mu^2/ (1-x)),$
the Feynman virtuality of the struck
parton\refmark\SS.
Furthermore, the integral of
the  $g_1$ structure function
is only affected
by QCD radiative corrections
of order $\alpha_s(Q^2)/\pi.$

\REF\Einhorn{%
M. B. Einhorn,
Phys. Rev. {\bf D14}, 3451 (1976).}

Thus PQCD can give useful predictions for the power law fall-off of
helicity-aligned and anti-aligned structure functions at $x \sim 1.$
Higher order contributions involving additional hard gluon exchange are
suppressed by powers of $\alpha_s(k_F^2)$.  Further iterations of the
interaction kernel will give factors of fractional powers of $\log
(1-x)$ analogous to the anomalous dimensions
$\log^{\gamma_n} Q^2$ which appear
in the PQCD treatment of form factors at large momentum
transfer\refmark\BLreview. This is in contrast
to super-renormalizable theories
such as QCD(1+1) where the power-law behavior
in the endpoint region is modified
by all-order contributions\refmark\Einhorn.

\REF\Hoyer{S.  J.  Brodsky, P.  Hoyer, A.  H.  Mueller and W.  Tang,
Nucl.  Phys. {\bf B369}, 519 (1992).}

\REF\Nagoya{S. J. Brodsky,
{\it Proceedings of the
10th International Symposium  on High-Energy
Spin Physics},
Nagoya, Japan  (1992).}

The fact that one has a definite prediction for the $x\sim 1$ behavior of
leading twist structure functions is a powerful tool in QCD
phenomenology, since any contribution that does not decrease sufficiently
fast at large $x$ is most likely due to coherent multi-quark
correlations.  As discussed in Ref.~[\Hoyer], such contributions are
higher twist, but they arise naturally in QCD and are significant at fixed
$(1-x)Q^2.$ Such coherent contributions are in fact needed in order to
explain the anomalous change in polarization seen in pion-induced
continuum lepton-pair and hadronic $J/\psi$ production experiments at
high $x_F$\refmark\Nagoya.

At large $x$ the perturbative QCD analysis predicts ``helicity retention"
-- \ie, the helicity of a valence quark
with $x \sim 1$ will matches that of the
parent nucleon. This result is in agreement with the original
prediction of Farrar and Jackson\refmark\Farrar~
that the helicity asymmetry $\Delta q(x)$
approaches 1 at $x \to 1.$ We also  predict, in agreement with
Ref. [\Farrar], that the ratio of unpolarized neutron to proton
structure functions approaches the value $3/7$ for $x \to 1$.

In the following sections we will analyze the shape of the polarized
gluon and quark distributions in the proton.  First we will study the
behavior of the gluon asymmetry $\Delta G(x)/G(x)$ (polarized over
unpolarized distributions) at small values of $x$, where it turns out to
be proportional to $x$ with a coefficient approximately independent of
the details of the bound-state wavefunction.  We then write down a simple
model for the gluon distributions which incorporates the counting rule
constraints at $x \to 1.$ The same is done for the up, down and strange
quark distributions.  The extrinsic and intrinsic strange quark
distributions are also discussed, paying special attention to the
inclusive-exclusive connection with the strange quark contribution to the
proton form factors.

{\bf \chapter {HELICITY-DEPENDENT GLUON DISTRIBUTIONS}}

\REF\JM{%
See, for example, R. L. Jaffe and  A. Manohar,
Nucl. Phys. {\bf  B337}, 509 (1990).}

\REF\MA{%
Bo-Qiang Ma,
Z. Phys. {\bf C58}, 479 (1993).}
The angular momentum
of a fast-moving  proton has three sources, the angular
momentum carried by the quarks,
the angular momentum carried by the gluons, and the orbital
angular momentum carried by any of the constituents.  Angular momentum
conservation for $J_z$ at a fixed light-cone time
implies the sum rule\refmark{\JM}~
$${1 \over 2} (\Delta u + \Delta d + \Delta s) +
\Delta G~+ <L_z> = {1 \over 2}. \eqn\Asum$$
Here $\Delta G \equiv \int_0^1 dx
\Delta G(x)$ is the helicity carried by the gluons, where $\Delta G(x)$
is the difference between the helicity-aligned and anti-aligned gluon
distributions $G^+(x)$ and $G^-(x)$; the unpolarized gluon distribution
$G(x)$ is the sum of these two functions, $G(x) \equiv G^+(x)+G^-(x)$.
The corresponding definitions for the quark distributions $\Delta q(x) =
q^+(x) - q^-(x)$ and $q(x)=q^+(x) + q^-(x)$ with $q=u,d,s.$ By
definition, the antiquark contributions are included in $\Delta q(x)$ and
$q(x).$ As emphasized by Ma\refmark\MA, the helicity distributions
measured on the light-cone are related by a Wigner rotation (Melosh
transformation) to the ordinary spins
$S^z_i$ of the quarks in an equal-time
rest-frame wavefunction description. Thus, due to the non-colinearity
of the quarks, one cannot expect that
the quark helicities will sum simply
to the proton spin.

In this paper we shall present model forms for the gluon distribution
functions $\Delta G(x)$ and $G(x)$ for nucleons which incorporate the
known large-$x$ counting-rule constraints:
$$G^+(x) \to C(1-x)^4 \quad\quad (x \to 1)\ ,\eqn\Aa $$
$$G^-(x) \to C(1-x)^6 \quad\quad (x \to 1)\ ; \eqn\Ab $$
We will also incorporate a constraint on
the behavior of the gluon asymmetry ratio $\Delta G(x)/G(x)$ for small
$x:$
$$\left({\Delta G(x) \over G(x)}\right)_{\rm proton} \to {x \over 3}
\VEV{{1\over y}} \quad\quad (x \to 0). \eqn\Ac $$
This last theoretical constraint will be demonstrated below.
Here $\VEV{1/y}$ stands
for the first inverse moment of the quark light-cone momentum fraction
distribution in the proton lowest Fock state.  For this state we expect
$\VEV{1/y} \simeq 3$.

A simple form for baryon gluon distributions, which incorporates the
limiting behaviors presented above, is
$$\eqalign {\Delta G(x) =& {N
\over x}\ [1-(1-x)^2]\ (1-x)^4 ,\cr
 G(x)=& {N \over x}\ [1+(1-x)^2]\ (1-x)^4 . } \eqn\Ad $$
In this model the momentum fraction carried by the gluons in the proton
is $\VEV{x_g}
\equiv \int^1_0 dx x G(x) ={12 \over 35} N$, and the helicity
carried by the gluons is $\Delta G \equiv \int^1_0 dx \Delta G(x) = (11/
30) N$.  Taking the momentum fraction $\VEV{x_g}$ to be $1/2$, we
predict $\Delta G = 0.54$.

Such large values for the gluon momentum fraction are
inconsistent with the assumption that the proton
has a dominant three-quark
Fock state probability; a  self-consistent approach
thus requires taking into account gluon radiation from
the full quark and gluon light-cone
Fock basis of the nucleon.
Our main emphasis here is to predict the characteristic
shapes of the polarized quark and gluon distributions.
The large $x$ regime is clearly  dominated by the
lowest particle-number Fock states.
We thus expect that the qualitative features of the model
to survive  in a
more rigorous approach; in particular, it is apparent from
the structure of the model,
that the gluon helicity fraction will be of the same
order of magnitude as the gluon momentum fraction.

The prediction that $\Delta G \simeq 0.5$ is phenomenologically
interesting.  If one also accepts
the experimental suggestion from EMC  that the quark
helicity sum $\Delta u + \Delta d$ is small, then this implies
that gluons could carry a significant fraction of the
proton helicity $J_z=1/2$ of the same size as the
momentum fraction carried by the gluons.
However, one  also expects significant
orbital angular momentum $L_z$ which arises,
for example, from the  finite transverse
momentum associated with the $q \to q
g$ gluon emission matrix element.

\REF\Stan{S.  J.  Brodsky and I. A. Schmidt, Phys. Lett. {\bf B234}, 144
(1990).  Coherence effects are also discussed in: S.  J.  Brodsky and J.
F.  Gunion, Phys.  Rev. {\bf D19}, 1005 (1979).  Applications to atomic
and molecular systems are discussed in:
M.  Burkardt, Nucl.  Phys. {\bf B373}, 371 (1992);
see also M.  Kalu\v za, A.~G.~%
Schneider-Neureither and H.  Pirner, Universit\"at Heidelberg preprint
(1993).}

\REF\QCDSUM{V. L. Chernyak and A. R. Zhitnitsky,
Phys. Rept. {\bf 112}, 173 (1984).}

We now proceed to prove equation~\Ac~for the low-$x$ behavior of the
asymmetry $\Delta G(x)/G(x)$.  In this region the quarks in the hadron
radiate coherently, and we must consider interference between amplitudes
in which gluons are emitted from different quark lines.  An analysis
of this type was first presented in Ref.~[\Stan], and in this note we
extend and correct some of the results of that paper.

As an example, we first analyze the helicity content of positronium, where
we can ignore internal transverse momenta and non-collinearity.
Consider the ortho-positronium two-fermion $J_z=1$ Fock state in which
the particles have helicities $+~+$.  Following the calculation of
Ref.~[\Stan], we obtain
$$\left({\Delta G(x) \over
G(x)}\right)_{{\rm ortho}(J_z=+1)} \simeq x
\VEV{{1 \over y}} \simeq 2 x \quad\quad
(x \to 0)\ . \eqn\Ae $$
In the case of para-positronium (and also for $J_z=0$ ortho-positronium),
in which we start with a Fock state with helicities $+~-$, the result is
$\Delta G(x) = 0$.  This is because for every diagram in $G^+(x)$ there
is a corresponding diagram in $G^-(x)$, but with the helicities of all
the particles reversed.

We now apply a similar analysis to the
gluon distribution in the nucleon.  We
start with a three-quark Fock state in
which the quarks have helicities $+~+~+$
as would be appropriate for the helicity
content of an isobar state $\Delta$
with $J_z=3/2$.  Then the result found in
Ref.~[\Stan], i.e. $$\left({\Delta G(x)
\over G(x)}\right)_{\Delta(J_z=3/2)} \simeq
x \VEV{{1 \over y}} \eqn\Af $$
follows.

In the nucleon case, however, we start with a three-quark Fock state with
helicities $+~+~-$.  Thus clearly there is a cancellation between the
squared terms in which the gluon is emitted from one of the positive
helicity quarks versus the contributions in which the gluon is emitted by
a negative helicity quark.  The interference terms work similarly,
ensuring a finite result for both $G(x)$ and $\Delta G(x)$ at zero
$k_{\perp},$ just as in the case of photon distributions in positronium.
Then the positive helicity quarks have a dominant $G^+(x)$ and
contribute positively to $\Delta G(x)$; similarly, the negative helicity
quarks contribute negatively to $\Delta G(x)$.  To see this more clearly,
consider the photon emitted by a single electron with $J_z=+1/2$.  Then
$G^+_{\gamma/e}(x)=1/x$
and $G^-_{\gamma/e}(x) =(1-x)^2/x$.  Thus $\Delta G(x)/G(x) = x $ at
$x \to 0$ with unit coefficient
in this case.  The sign reverses for an electron with
$J_z=-1/2$.

The generated gluon asymmetry distribution in the nucleon at low $x$ is
then given by equation~\Ac.  The extra factor of $1/3$ is due to the fact
that all the quarks contribute positively to $G(x)$, but they give
contributions proportional to the sign of their helicity in $\Delta
G(x)$.  The main assumption setting the value of the gluon asymmetry
at $x \to 0$ is the  estimated
value of the inverse moment $\VEV{1/y}$.  For realistic wavefunctions
this expectation value may receive very large (possibly divergent)
contributions from near $y=0$.  However, one must be careful at this
point because in deriving Eq.~\Ac~
we assumed that $x \ll y$.  In order to
be consistent with this assumption we will perturb around a constituent
quark wavefunction which is strongly peaked around
$y=\VEV{y}=1/3$.  We have
furthermore assumed for simplicity that
$\VEV{y}$ is the same for all valence
quarks, although this is inconsistent with results from $QCD$ sum rules
\refmark\QCDSUM.
(One could improve the estimate for $\VEV{1/y}$ by allowing
for different momentum fractions for
the helicity-up and helicity-down quarks.
This would evidently reduce $\Delta G$,
since it is known that $\VEV{y}$ is larger
for helicity parallel quarks.
Furthermore, in QCD we expect that higher Fock
states will contribute to reduce the value of
$\VEV{y}$ away from $1/3$,
which would be the expected value if only the three-quark valence Fock
state was present.)

{\bf \chapter {HELICITY-DEPENDENT QUARK DISTRIBUTIONS}}

In this section we shall construct a simple polynomial model
for the helicity-dependent
quark distributions in the proton and neutron.

As we have discussed in the previous sections,
at $x \sim 1 $ PQCD predicts
that the helicity-parallel quark distribution $q^+(x)$ is enhanced
relative to the helicity-antiparallel quark distribution $q^-(x)$ by two
powers of $(1-x)$.  The property of helicity retention at large $x$ is a
direct consequence of the gauge theory couplings between quarks and
gluons.
For the valence quarks in a nucleon the counting rules predict
$$q^+(x) \sim (1-x)^3 \quad\quad (x \to 1), \eqn\Ba $$ and
$$q^-(x) \sim (1-x)^5 \quad\quad (x \to 1). \eqn\Bb $$
The case of the non-valence strange quarks is somewhat
more complex and will be discussed in detail in the next section.  The
result is
$$s^+(x) \sim (1-x)^5 \quad\quad (x \to 1), \eqn\Bc $$
$$s^-(x) \sim (1-x)^7 \quad\quad (x \to 1). \eqn\Bd $$
For $x \sim 0$ the helicity correlation disappears since the
constituent has infinite rapidity $\Delta y = \log x$ relative to the
nucleon's rapidity.

The strange quark distribution in a nucleon can arise from both
intrinsic and extrinsic contributions.
The intrinsic contribution is associated with
the multiparticle Fock state decomposition
of the hadronic wavefunction, and it is essentially of
non-perturbative origin. This is in contrast to the extrinsic component,
which
arises from $s \bar s$ pair production from a gluon emitted by a valence
quark, and is associated with the self-field
of a single quark in the proton.
From evolution and gluon splitting, the extrinsic strange contributions
are known to behave as
$$s_e^+(x) \sim (1-x)^5    \quad\quad (x \to 1) \eqn\Bj $$
$$s_e^-(x) \sim (1-x)^7    \quad\quad (x \to 1)\ . \eqn\Bk $$

The Drell-Yan inclusive-exclusive connection relates the high $Q^2$
behavior of the hadronic form factors to the large $x$ limit of the
quark distribution functions; i.e.
$$F(Q^2)~ {\buildrel Q^2 \to \infty \over \longrightarrow}~
{1 \over (Q^2)^n}\quad     \Longleftrightarrow   \quad
G_{q/p}~ {\buildrel x \to 1 \over \longrightarrow} ~
(1-x)^{2n-1+2\Delta S_z},                         \eqn\Bl   $$
where $\Delta S_z = 0$ or $1$ for parallel or antiparallel quark
and proton helicities, respectively. If we naively apply this
prescription to the extrinsic strange quark component, we would
predict  that the strange quark contribution to the electromagnetic
proton form factor should fall as $1/Q^6$, since in this case $n=3$.
But a direct calculation of the strange quark contribution to either
the axial or vector form factor of the nucleon
gives only a nominal $1/Q^4$ behavior, which is the
same power-law fall-off as the valence quark contribution.
In the leading order calculations
the loop integrals connecting a hard $s \bar s$  loop
to a valence quark
all have momenta $\ell = {\cal O} (Q),$ thus producing
radiative corrections of order $\alpha^N_s(Q),$ to the exclusive
amplitude
with $N=2$ (axial) or $N=3$ (vector),
rather
than extra powers of $1/Q^2$\refmark\LB.
The solution to this apparent contradiction
is that we should apply
the inclusive-exclusive connection for the strange quark contributions
to a transition form factor connecting
an initial state with three quarks ($uud$) to a final state in which
a strange pair has been created ($uuds\bar s$), as in the
transition form factor $p \to \Lambda K,$ at {\it fixed} final state
mass. Since the internal hard-scattering matrix element $T_H$ for
$(uud) + \gamma^* \to sudu \bar s$ has three off-shell fermion
legs, this transition form factor falls off as $(1/Q^2)^3,$ and
it correctly satisfies the inclusive-exclusive connection ($n=3$).

One can also consider the case where $Q^2$ and the final state mass
are both large, but there is a $K$ and $\Lambda$ in the final state.
This again corresponds to a $\sim (1-x)^5$ structure function.
In the case of the transition $p \to p \phi$, there is a color
mismatch in $T_H$ at lowest order. Thus this amplitude should be
suppressed (Zweig rule) by an extra power of $\alpha_s(Q^2)$.
Of course all of this holds for the analogous charm systems as well.

The intrinsic strange components are associated with Fock states
having at least five particles; the distributions thus have the
behavior $$s_i^+(x) \sim (1-x)^7   \quad\quad (x \to 1)  \eqn\Bm  $$
$$s_i^-(x) \sim (1-x)^9   \quad\quad (x \to 1)\ ,  \eqn\Bn$$
which corresponds to $n=4$ in the spectator
quark counting rules. It also satisfies the inclusive-exclusive
connection, since the intrinsic contribution to the form factor
falls as $(1/Q^2)^4$.

For the complete parameterization we shall adopt the canonical forms:
$$u^+(x) = {1\over x^\alpha} \left[A_u (1-x)^3 + B_u (1-x)^4 \right],$$
$$d^+(x) = {1\over x^\alpha} \left[A_d (1-x)^3 + B_d (1-x)^4 \right],$$
$$u^-(x) = {1\over x^\alpha} \left[C_u (1-x)^5 + D_u (1-x)^6 \right],$$
$$d^-(x) = {1\over x^\alpha} \left[C_d (1-x)^5 + D_d (1-x)^6 \right],$$
$$s^+(x) = {1\over x^\alpha} \left[A_s (1-x)^5 + B_s (1-x)^6 \right],$$
$$s^-(x) = {1\over x^\alpha} \left[C_s (1-x)^7 + D_s (1-x)^8 \right],$$
where we require
$$A_q+B_q=C_q+D_q$$
to ensure the convergence of the helicity-dependent sum rules.
Thus in our model, the Regge behavior of the asymmetry $\Delta q(x) \sim
x^{-\alpha_R}$ is automatically one unit less than the unpolarized
intercept: $\alpha_R = \alpha-1.$
Isospin symmetry at low $x$ (Pomeron dominance) also requires
$$A_u+B_u+C_u+D_u=A_d+B_d+C_d+D_d\ .$$
We emphasize that these distributions
include both the quark and antiquark contributions.

\REF\MRS{%
A. D. Martin, W. J. Stirling, and  R.G. Roberts,
Phys. Lett. {\bf B306}, 145 (1993),
RAL-94-055,
HEP-PH-9406315.}

\REF\HERA{%
ZEUS Collaboration, Phys.Lett. {\bf B315}, 481 (1993).}

Our parameterization of the
helicity-dependent quark distributions is close in
spirit to the  parameterization  $D_0'$ for the
unpolarized quark and gluon distributions given by Martin, Roberts and
Stirling\refmark\MRS. The MRS parameterizations are a good match to our
unpolarized forms $q(x)= q^+(x)+q^-(x)$ since the MRS forms combine
counting-rule constraints
with a good fit to a wide range of perturbative QCD
phenomenology.
We find that choosing the
effective QCD Pomeron intercept $\alpha = 1.12$ allows good match to
the unpolarized quark distributions given by the
MRS parameterization $D_0'$ at $Q^2 = 4$ GeV$^2$
over the range $0.001 < x < 1.$
It also predicts an increasing
structure function $F_2(x,Q^2)$ for
$x < 10^{-3},$ as suggested
the recent data from HERA\refmark\HERA.
Thus  we predict
$\alpha_R = 0.12$ for
the helicity-changing Reggeon intercept.  The momentum
fraction carried by the
quarks (and antiquarks),  $\VEV{x_q}= \int_0^1 dx x q(x)$,
where $q(x) \equiv q^+(x) + q^-(x)$, is assumed to be $ \sim 0.5.$

\REF\SLACa{M. J. Alguard \etal, Phys. Rev.
Lett. {\bf 37}, 1261 (1976); {\bf
41}, 70 (1978); G. Baum et al., ibid {\bf 51}, 1135 (1983).}

\REF\EMC{J. Ashman  \etal,
Phys. Lett. {\bf B206}, 364 (1988); Nucl. Phys. {\bf
B328}, 1 (1989).}


\REF\BEK{See for example, S. J. Brodsky, J. Ellis, and M. Karliner,
Phys. Lett. {\bf 206B}, 309, (1988).}

\REF\Schlumpf {%
S. J. Brodsky and F. Schlumpf
Phys. Lett. {\bf 329B}, 111, (1994).}

A combined analysis\refmark\Jaffe~ of the
SLAC and EMC\refmark{\SLACa,\EMC}\  polarized electron-proton
data
provides the constraint:
$$\int dx g^p_1(x) = 0.112\pm 0.009\pm 0.019\ .$$
If one uses the central value
together with the constraints from nucleon and hyperon decay and
includes radiative corrections of ${\cal O}((\alpha_s/\pi)^3)$
then one obtains the
following values for the proton helicity carried by the different
quarks\refmark\Jaffe:
$$\Delta u = 0.83\pm 0.03, \quad \Delta d =-0.43\pm 0.03, \quad
\Delta s = -0.10\pm 0.03\ , \eqn\Be $$
at the renormalization scale $Q^2=10\ {\rm GeV}^2$.
Since these values for the $\Delta q$ are  obtained after removing the
deep inelastic radiative corrections, we can
use them
as the initial
phenomenological inputs for the proton; the
neutron distributions then follow from isospin symmetry.
The small value for the total quark helicity
$\Delta \Sigma = \Delta u+ \Delta d + \Delta s = 0.31 \pm 0.07$
is consistent with
large $N_C$ predictions in QCD\refmark\BEK,
and it is about half of the value $\Delta \Sigma \simeq 0.75$
expected in
the relativistic three-quark constituent model
for the nucleon without dynamical gluons\refmark\Schlumpf.
As we shall show below, the gluon helicity fraction
$\Delta g$ scales
closely with the gluon momentum fraction $\VEV {x_g}.$

\REF\NMC {%
NMC Collaboration,  Phys. Rev. D{\bf 50} 1,(1994).
P. Amaudruz, \etal, Phys. Rev. Lett. {\bf 66} 2712 (1991).}

The $u(x)$ and $d(x)$ parameterizations have eight
parameters which we will fix
using the following eight conditions:
three conditions arise from the requirement
that the sum rules converge
at $x \to 0$; two conditions come from
the values of $\Delta u$ and $\Delta d$; one condition follows
by imposing the $SU(6)$ large $x$ relation
$A_u = 5 A_d$; one condition
is obtained from the empirical value of the
Gottfried sum $S_G\equiv \int[u(x)-d(x)]/3=0.235$\refmark\NMC;
the final condition is obtained from the sum of
momentum fractions carried by the quark and antiquark
$x_u+x_d=0.521$\refmark\MRS.
It is straightforward to find
parameters for the polynomial forms which are
consistent with the above inputs.
$$  A_u = 3.784, \quad   A_d = 0.757,  \eqn\Bf  $$
$$  B_u =-3.672, \quad   B_d =-0.645,  \eqn\Bg  $$
$$  C_u = 2.004, \quad   C_d = 3.230,  \eqn\Bh  $$
$$  D_u =-1.892, \quad   D_d =-3.118,  . \eqn\Bi  $$
With this set of parameters, the respective quark momentum
fractions are:
$$\VEV{x_u}  = 0.331,
\quad \VEV{x_d}  = 0.190 \eqn\Bk $$
The predicted distributions $xu(x), xd(x), \Delta d(x),$
and $\Delta u(x)$  are shown in Figs. 1(a) and 1(b).
In each case both the quark and antiquark contributions
are included.   The simple polynomial forms
represent a simple parameterization of
the non-perturbative polarized and unpolarized
quark distributions which satisfy the known
theoretical constraints at large and small $x$ and
the empirical sum rules.
We also show a comparison of the unpolarized
distributions with
the MRS $D_0'$  phenomenological
parameterizations.  The agreement is quite reasonable.
The differences in the shapes of the distributions
can be attributed to the effects of perturbative QCD evolution.

Notice that
$\Delta d(x) \equiv d^+(x)-d^-(x)$ is positive at
large $x$ (which follows from $A_u=5 A_d$),
and negative at small to moderate values of $x$.
We thus  predict
that $\Delta d(x)$ will change sign and  go through zero at some
physical value for $x$. With the above parameterization the
zero of $\Delta d(x)$ occurs at $x=0.489.$

\REF\CCFR{%
CCFR, Phys. Rev. Lett {\bf 70}, 134 (1993).}.

\REF\prep{%
For an alternative parameterization of the strange quark
distributions, see G. Preparata, P. G. Ratcliffe,
and J. Soffer,
Phys. Lett. {\bf B273} 306, (1991).}

In the case of the strange quark plus strange antiquark
distributions,
we have four
parameters and three conditions: one from the convergence of sum rules;
one from the value of $\Delta s$; and one from the momentum
fraction carried
by strange plus anti-strange quarks $x_s = 0.035$\refmark\CCFR.
This leaves us with one unknown, which we choose to be $C_s.$
The three constraints give the solution
set:
$$A_s= -0.6980 + 0.9877 C_s, \quad B_s = 0.8534-1.1171C_s,
\quad D_s = 0.1551-1.1294C_s
\eqn\Sa $$
Because of the probabilistic interpretation of
parton distribution functions, $s^+$(x) and $s^-(x)$ both
must be non-negative functions of $x,$ which implies the rather
stringent bounds
$$ 0.7067 < C_s < 1.2013 .$$
Within these bounds, $g_1(x)$ is practically independent
of $C_s$;  to be definite, we chose $C_s=1.$
(We could have taken any other value consistent with the
inequalities\refmark\prep.)
We compare our simple parameterization to the MRS
$D_0'$ parameterization
in Fig. 1(c). The MRS distribution which gives an
approximate realization of the data rises faster at
low $x$ than our model. This could be attributed to
the need to impose higher Pomeron intercept, or the
effects of QCD evolution.

We can also find
parameterizations for the polarized
gluon distributions which
are consistent with the $x \to 0$ and $x \to 1$ helicity constraints,
as well as the MRS unpolarized gluon distribution:
$$G^+(x) = {1 \over x^{\alpha_g}}
\left[A_g (1-x)^4 + B_g (1-x)^5 \right]\ , \eqn\Bl$$
$$G^-(x) = {1 \over x^{\alpha_g}}
\left[A_g (1-x)^6 + B_g (1-x)^7 \right]\ , \eqn\Bm$$
This form automatically incorporates
the coherence constraint,  Eq. \Ac.
We shall assume that $\alpha_g=\alpha=1.12$ so that
the pomeron intercept is identical for quark and
gluon distributions.  The parameter set
$A_g = 2$ and  $B_g = -1.25 $
gives an unpolarized gluon distribution
$G(x)=G^+(x)+G^-(x)$
similar to the
phenomenological $D_0'$ gluon
distribution given by MRS.  (See Fig. 2.)
The momentum carried by the gluons in the nucleon
using the above simple form is $\VEV{x_g}=0.42.$
(The gluon and light quark and antiquark
distributions then almost saturate the
momentum sum rule.)
The
gluon helicity content for the above parameterization
is $\Delta G = 0.45.$As shown in the figure,
the shape of the polarized distribution $\Delta G(x)$
given by the above parameterization
is almost identical to $xG(x)$.

Alternatively, if we take
$\alpha_g= 1$, then the parameter set
$A_g = 0.2381$ and   $B_g = 1.1739.$
give again the same values $\VEV{x-g}=0.42$ and
$\Delta G=0.45$ as above.
In this case,
the resulting shape
unpolarized  distribution
$G(x)=G^+(x)+G^-(x)$
is indistinguishable from the
phenomenological $D_0'$ gluon
distribution given by MRS.

\REF\Morii{%
T. Morii, S. Tanaka and T. Yamanishi, Phys. Lett. {\bf B322},
253 (1994). See also:
A. V. Efremov, L. Mankiewicz and N. A. Tornqvist, Phys. Lett.
{\bf B284}, 394 (1992).}

Although there is some experimental information about the unpolarized
gluon distribution, this is not the case for the polarized
gluon distribution. It is important to test these
distributions  directly, for example  in processes such as
$J/\psi$ production in polarized $e-p$ and
$p-p$ collisions\refmark\Morii.

{\bf \chapter {POLARIZED STRUCTURE FUNCTIONS}}

In this section
we will use the polynomial model forms
for $\Delta q(x)$ and $q(x)$ to
compute the polarized helicity structure functions of nucleons:
$$g_1^{ep}(x) = {1 \over 2}\left( {4\over 9}\Delta u(x)
  +{1\over 9}\Delta d(x)   + {1\over 9}\Delta s(x) \right) \eqn\Ca $$
and
$$g_1^{en}(x) = {1 \over 2}\left( {4\over 9}\Delta d(x)
  +{1\over 9}\Delta u(x)   + {1\over 9}\Delta s(x) \right)\ ,\eqn\Cb $$
and compare the results to the recent experiments.
(Note that $\Delta q(x)$ refers to the
combined asymmetries from both quarks and
antiquarks in the proton.) A precise prediction requires
the inclusion of QCD evolution. Here
we will, as in  Ref. [\Jaffe], simply include
the normalization factor $N_{QCD} = 1 - (\alpha_s/\pi) \approx .92$
arising from $QCD$ radiative corrections.
The Bjorken sum rule for the difference of proton and
neutron quark helicities is automatically satisfied.  The
Ellis-Jaffe sum rule for
the nucleon quark helicity is violated by the model due
to the presence of the strange quark contributions $\Delta s.$

\REF\Gibson{See for example: W. M. Gibson and B. R. Pollard, in
{\it Symmetry Principles in Elementary Particle Physics,}  Cambridge
University Press (1976), page 330.}

We have emphasized that the dynamics of QCD implies helicity retention:
the quark with $x$ close to $1$ has the same helicity as the proton.
Thus all of the structure
functions asymmetries become maximal at $x \to 1$, and  the ratio of
unpolarized proton and neutron structure functions can be predicted.

According to the standard $SU(6)$ flavor and helicity symmetry, the
probabilities to find $u$ and $d$ quarks of
different helicities in the proton's
three-quark wavefunction are: $P(u^+)=5/9, P(d^+)=1/9, P(u^-)=1/9,
P(d^-)=2/9$\refmark\Gibson. Thus the usual expectation from SU(6)
symmetry is $F_2(n)/F_2(p)=2/3$ for all $x$.  As Farrar and Jackson
pointed out\refmark\Farrar, this naive $SU(6)$ result
cannot apply to the local
helicity distributions since the helicity aligned and helicity
anti-aligned distributions have different momentum distributions.
At large $x$
$u^-$ and $d^-$ can be neglected relative to $u^+$ and $d^+$,
and thus $SU(6)$
is broken to $SU(3)^+ \times SU(3)^-$.  Our model retains the
$SU(6)$ ratio $P(u^+)):P(d^+))= A_u:A_d
= 5:1,$ at large $x$ so that we predict
$F_2(n)/F_2(p) \to 3/7$ as $x \to 1.$
The physical picture that emerges is that
the struck quark carries all the helicity
of the nucleon, and the spectators have
$S_z=0$, although their total helicity is
a combination of $0$ and $1$.  This
wavefunction is just a piece of the full $SU(6)$ wavefunction, but since
it is the piece that contains the $u^+$ and $d^+$, and this part remains
unchanged, the ratio $P(u^+)/P(d^+)$ is still $5/1.$

\REF\CK{R.  D.  Carlitz and J.  Kaur, Phys.  Rev.  Lett. {\bf 38}, 673
(1977); J.  Kaur, Nucl.  Phys.  {\bf B128},
219 (1977).  For recent empirical
models, see Ref. [\BergerQiu]~
and K.~Kobayakawa  \etal, Phys.  Rev. {\bf
D46}, 2854 (1992).}

Notice that the only empirical input into our model is the integrated
values of the various flavors obtained from the proton data.  The shape
of the polarized distributions
is essentially
determined by the perturbative QCD constraints.  The agreement with the
shape of the SLAC and EMC experimental data for the proton is rather
good (see Fig. 3(a)) and could be further improved by taking into
account PQCD evolution.

We can also
compare our model with the polarized neutron structure function
extracted by the E142 from its polarized electron polarized
$He^3$ measurements. (See Fig. 3(b).)
For the neutron we predict two new effects
which are not present in the proton.  First $g_1^{en}$ tends to fall
faster than $g_1^{ep}$ for large x.  This is because as in the
Carlitz-Kaur\refmark\CK~ and
Farrar-Jackson\refmark\Farrar~ models, the helicity
aligned up-quark dominates the proton distribution and the helicity down
quark dominates the neutron structure function at large $x.$ A related
effect is that $g_1^{en}(x)$ changes sign as a function of $x$.  This is
due to the fact that except for large $x$ (where the helicity aligned
down quark dominates) $g_1^{en}$ is dominated by the anti-aligned up
quark distribution.  Since $\int_0^1dx\Delta u_n(x) = \int_0^1 dx\Delta
d_p(x) < 0$\refmark\Jaffe~ it is
clear that $g_1^{en}(x)$ must be negative
at small $x.$

A comparison of our model with the recent SMC data for the polarized
deuteron structure function $g^d_1(x)$ is
shown in Fig. 4.  The shape of the data
appears to be consistent with our
predictions, except possibly at the largest $x$ point where the model
predicts too little asymmetry.   To make this prediction we have,
as in Ref. [\SMC],
assumed that the deuteron structure function is half of the
sum of the neutron and proton structure functions and included the
$D$-state depolarization factor with $D$-state probability 0.058.
The model then predicts the normalization
$$\eqalign{\int dx g_1^d(x) &= \half \int dx (g^p_1(x) + g^n_1(x))
\cr\crr
&= \left[{5\over 36}(\Delta u + \Delta d) + {1 \over 18}\Delta s\right]
\left(1-{\alpha_s\over\pi}\right)\ \left(1-{3\over 2} \omega_D\right)
= 0.038\cr}$$
compared to the
SMC result
$$\int dx g_1^d(x)=0.023
\pm 0.020 {\rm (stat.)}\pm 0.015 {\rm (syst.)}\ .$$

The distributions presented in this paper have applicability to any
PQCD leading-twist processes which require polarized quark and gluon
distributions as input.    Our input
parameters have been adjusted to be compatible with global parameters
available current experiments.
The values can be refined  as further and more
precise polarization experiments become available.
A more precise parameterization should also take into
account corrections from QCD evolution,
although this effect is relatively
unimportant for helicity-dependent distributions.
Our central observation is that
the shape of the distributions is almost completely predicted
when one employs the constraints
obtained from general QCD arguments at  large
$x$ and small $x.$

A remarkable prediction of our
formalism is the very strong  correlations between the parent hadron
helicity and each of its valence-quark, sea-quark,
and gluon constituents at
large light-cone momentum fraction $x$.  Although the
total quark helicity content of the proton is small,
we predict a strong positive
correlation of the proton's helicity
with the helicity of its $u$ quarks and
gluon constituents.
The model is also consistent
with the assumption that the
strange plus anti-strange quarks carry $3.5\%$ of the proton's momentum
and $-10\%$ of its helicity.
We also note that completely
independent predictions based on QCD sum
rules also imply that the three-valence-quark
light-cone distribution amplitude
has a very strong positive correlation at large $x$
when the $u-$quark and proton
helicities are parallel\refmark\QCDSUM.

 \endpage

\centerline{\bf ACKNOWLEDGEMENTS}

We would like to thank Peter Lepage, Al Mueller,
Michael Peskin, Felix Schlumpf,
and Wai-Keung Tang for helpful conversations.

\centerline{\bf FIGURE CAPTIONS}

Fig. 1:
Model predictions for the non-perturbative
polarized $\Delta q(x)=q^+(x)-q^-(x)$
and
unpolarized quark  $x q(x) = x(q^+(x)+q^-(x))$ distributions
in the proton.
The polynomial forms satisfy sum rule and dynamical constraints.
The leading Regge behavior at $x \to 0$ has intercept
$\alpha = 1.12.$
By definition both quark and antiquark contributions are included.
Comparison with the MRS $D_0'$ parameterization for the unpolarized
quark distributions\refmark\MRS\
are also shown.
(a): $u(x)$ distributions. (b) $d(x)$ distributions. (c) $s(x)$
distributions.

Fig. 2:
Predictions for the non-perturbative
polarized $\Delta G(x)=G^+(x)-G^-(x)$
and
unpolarized gluon  $x G(x) = x(G^+(x)+G^-(x))$ distributions
in the proton.
The polynomial forms satisfy sum rule and dynamical constraints.
The leading Regge behavior at $x \to 0$ has intercept
$\alpha_g = 1.12.$
Comparison with the MRS $D_0'$ parameterization for the unpolarized
gluon distributions\refmark\MRS\
is also shown.

Fig. 3(a):
Model prediction for the
polarized  helicity structure function of the
proton  compared with experiment.
Full line: sum of all flavors; dashed: only up quarks;
dotted: only down quarks; dash-dotted: only strange quarks.
We have multiplied our prediction with a PQCD correction factor
$1-(\alpha_s/\pi)=0.92$.
The data are from SLAC-EMC (closed circles), EMC (closed
squares), SMC (open squares) and
SLAC-E143 (diamonds).
Fig. 3(b):
same as (a) but for the neutron.
The data are from the SLAC E142 experiment\refmark\Emlyn.

Fig. 4:
Polarized  helicity structure function of the deuteron.
The data are from Ref. [\SMC]. We have multiplied our prediction
from the sum of proton and neutron contributions by
a $D-$state depolarization factor $1-(3/2) \omega_D$ with $\omega_D
=0.058$
and the PQCD correction factor $1-(\alpha_s/\pi)=0.92.$

\refout

\end

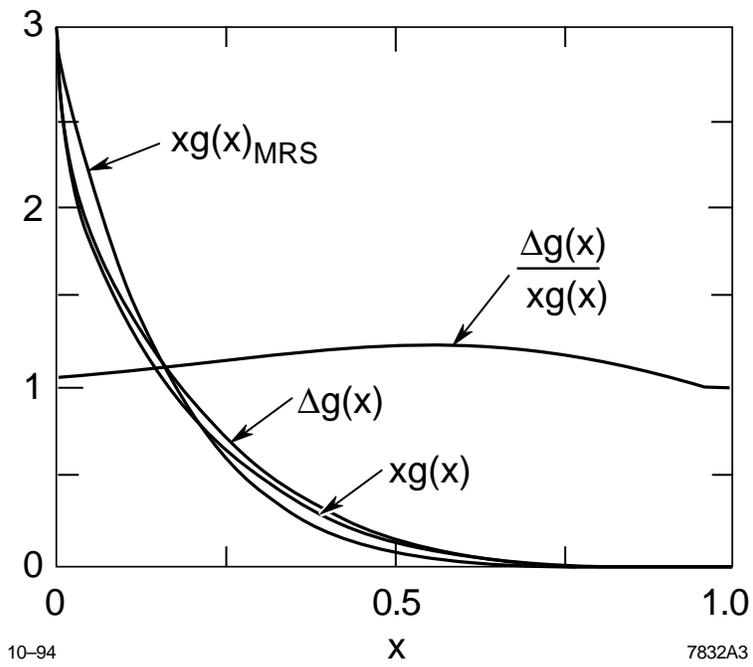
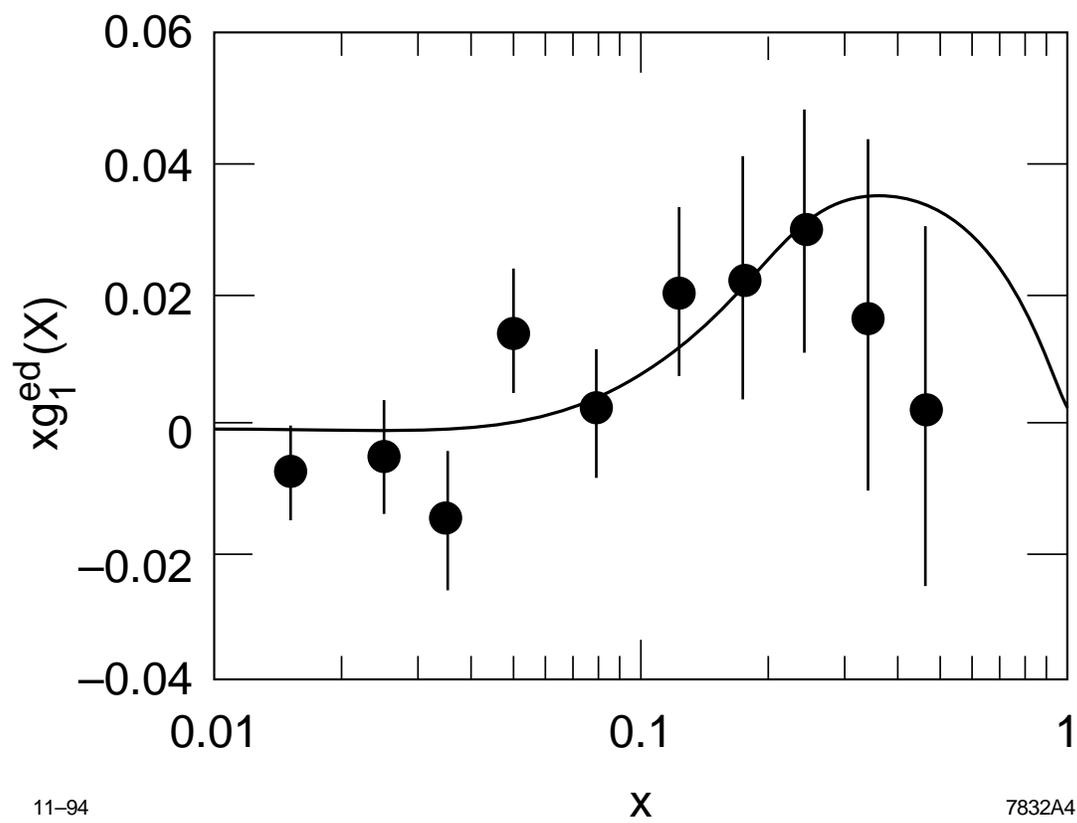
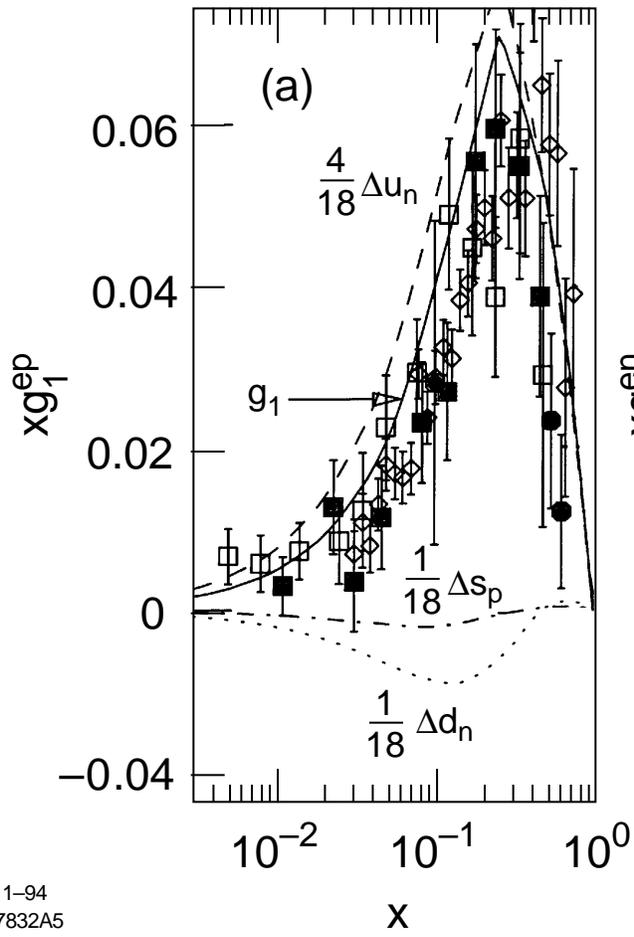
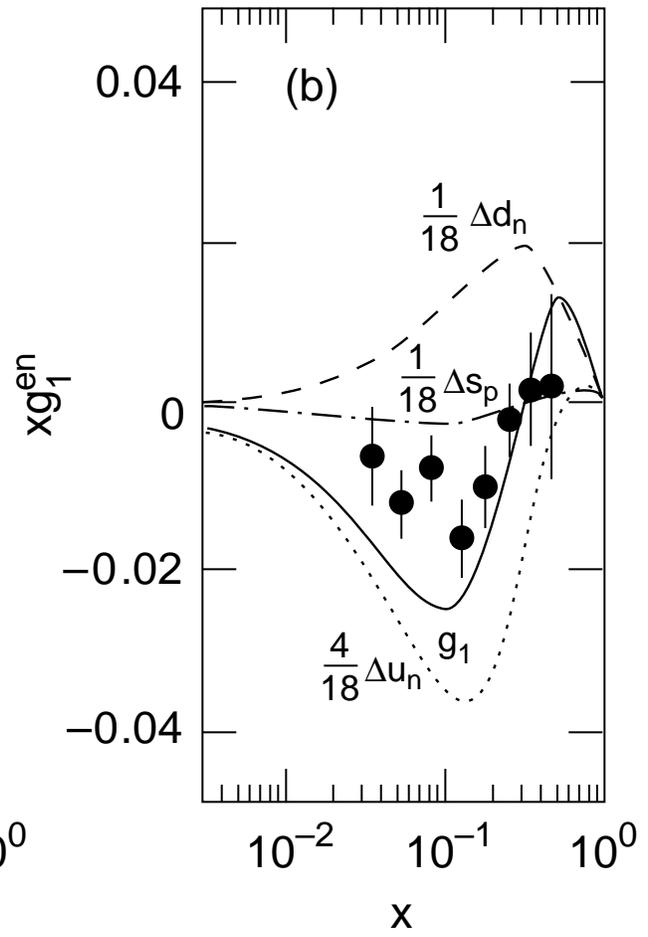
\end